\documentstyle[aps,epsf,prl,twocolumn]{revtex}
\begin{document}

\title{Lambda-proton correlations in relativistic heavy ion collisions}
\author{Fuqiang Wang$^1$ and Scott Pratt$^2$}
\address{
$^1$ Nuclear Science Division, Lawrence Berkeley National Laboratory, 
Berkeley, CA~94720, U.S.A.\\
$^2$ Department of Physics and National Superconducting Cyclotron Laboratory,
Michigan State University, East Lansing, MI~48824, U.S.A.}

\maketitle

\begin{abstract}
The prospect of using $\Lambda p$ correlations to extract source sizes
in relativistic heavy ion collisions is investigated. 
It is found that the strong interaction induces a large peak in the 
correlation function that provides more sensitive source size measurements
than $pp$ correlations under some circumstances.
The prospect of using $\Lambda p$ correlations to measure the time lag 
between lambda and proton emissions is also studied.
\end{abstract}



Two-particle correlations have proven to be a powerful tool for 
determining source sizes and lifetimes in heavy ion collisions.
At low energies, correlations of protons, neutrons and 
intermediate mass fragments have provided information on the 
space-time extent of the collision systems~\cite{low_energy}.
At relativistic energies, pion, kaon and proton correlations 
have greatly enhanced our understanding of the dynamics of 
heavy-ion collisions~\cite{boson_hbt};
these correlations provide different but complementary information.
For instance, heavier particles are more affected by collective flow, 
thus making the mass-dependence of source sizes a test of our picture 
of explosive flow in heavy ion collisions;
freeze-out conditions may be different for pions, kaons and protons,
thus comparing parameters inferred from their correlations allows one 
to test the conjecture of sequential freeze-out.

In this letter, we explore lambda-proton ($\Lambda p$) correlations 
as a candidate of interferometric study. 
We find that an enhancement to the correlation function at low 
relative momentum allows one to infer the size of the emitting source.
The inferred lambda source parameters may provide valuable information
because lambdas are strangeness carrying baryons.
Unlike two-proton ($pp$) system, the $\Lambda p$ system 
has no repulsive Coulomb interaction. 
Thus the enhancement from the strong interaction better survives 
when source sizes become large. We illustrate the sensitivity of
$\Lambda p$ correlations and show that for large sources, 
they might be more sensitive than $pp$ correlations, 
but not as sensitive as coalescence measurements. 
We also study the possibility to determine whether lambdas and protons
are emitted simultaneously by comparing the correlations for positive
and negative values of the projected outward relative momentum.

The correlation of two particles from a chaotic source may be estimated 
by assuming that they interact only with each other after they are 
emitted from space time points $x_a$ and $x_b$~\cite{gelbke},
\begin{eqnarray}
C({\bf p}_a,{\bf p}_b) &=& \frac{{\cal P}({\bf p}_a,{\bf p}_b)}
	{{\cal P}({\bf p}_a){\cal P}({\bf p}_b)} \nonumber \\
 & & \hspace*{-40pt} \approx 
\frac{\int d^4x_a d^4x_b S_a(p_a,x_a) S_b(p_b,x_b)
      |\phi_{\rm rel}({\bf p}_b-{\bf p}_a)|^2}
     {\int d^4x_a d^4x_b S_a({\bf p}_a,x_a) S_b({\bf p}_b,x_b)}
\end{eqnarray}
In principle, the correlation depends on the size and shape of 
the source described by function $S({\bf p},x)$, which provides 
the differential probability of emitting particles of 
momentum {\bf p} at a space-time point $x$. 
However, for the purposes of our study we will ignore the momentum
dependence of the source functions and assume a Gaussian form for $S$.
\begin{equation}
S(x_a) = \delta(t) \exp{\left( -\frac{x^2+y^2+z^2}{2R_g^2} \right)},
\end{equation}
where $R_g$ is the Gaussian size of the source.

The correlation function's sensitivity to the source size depends on
the form of the relative wave function, $\phi_{\rm rel}$. 
The relative wave function is determined by the relative strong 
interaction, the relative Coulomb interaction, 
and in the case of identical particles, symmeterization constraints. 
Since the $\Lambda p$ system involves non-identical particles, 
and interacts mostly through the $s$-wave channel at low relative momentum,
it is largely insensitive to details about the shape of the emitting source. 
However, since the strong interaction is short range, the enhancement of the 
correlation function at low relative momentum is highly sensitive to the size. 
Furthermore, the lack of a Coulomb repulsion, which dominates $pp$ 
correlations at low relative momentum, allows the $\Lambda p$ correlation
function to remain sensitive to the volume for fairly large sources.

An Urbana-type potential which was motivated from low energy $\Lambda p$
scattering and hypernuclei bind energy data~\cite{urbana} is used to 
generate the relative wave functions.
\begin{equation}
\label{eq:urbana}
V_{\Lambda p} = V_C - \left( \bar{V} - \frac{1}{4} V_{\sigma} 
\mbox{\boldmath$\sigma$}_{\Lambda} \cdot \mbox{\boldmath$\sigma$}_p
\right) T_\pi^2,
\end{equation}
where $V_C$ is a Woods-Saxon repulsive core.
\begin{equation}
V_C = W_C \left[ 1 + \exp\left( \frac{r-R}{d} \right) \right]^{-1},
\end{equation}
with $W_C$=2137 MeV, $R$=0.5 fm, $d$=0.2 fm. 
The modified one-pion exchange tensor potential
\begin{equation}
T_\pi = \left( 1+\frac{3}{x}+\frac{3}{x^2} \right) 
	\frac{e^{-x}}{x} \left( 1-e^{-cr^2} \right)^2,
\end{equation}
where $x$=0.7$r$ and $c$=2 fm$^{-2}$. 
The spin-independent part of the attractive potential is characterized
by $\bar{V} = 6.2 \pm 0.05$~MeV, while the spin-dependent part is small, 
$V_{\sigma} = 0.25 \pm 0.25$~MeV, and not well determined.

\begin{figure}
\centerline{\epsfxsize=0.45\textwidth \epsfbox{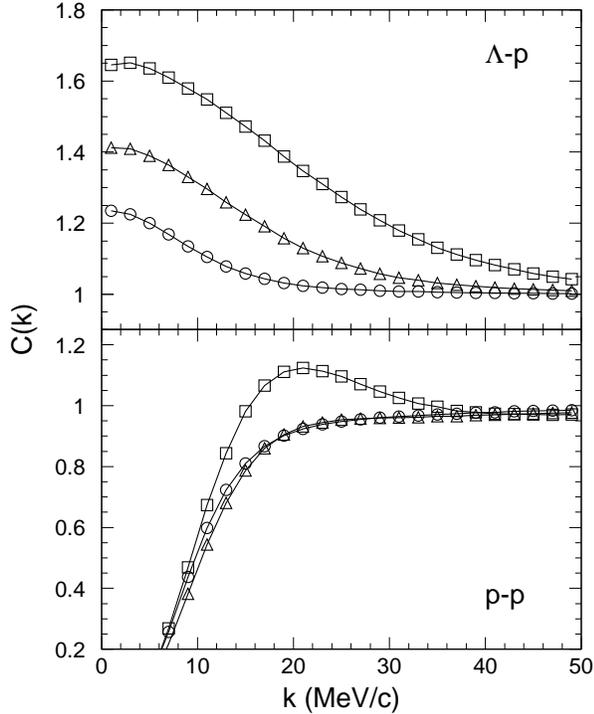}}
\caption{$\Lambda p$ (upper panel) and $pp$ (lower panel) correlation
functions for $R_g$=4~fm (squares), 6~fm (triangles) and 10~fm (circles).
For larger sources, $\Lambda p$ correlations provide a more sensitive 
determination of the size.}
\label{fig1_size}
\end{figure}

The correlation functions for $R_g = 4,6$ and 10 fm sources are 
illustrated in Fig.~\ref{fig1_size} as a function of $k$, 
which is one half of the relative momentum, 
$k=|{\bf p}_{\Lambda}-{\bf p}_p|/2$,
as measured in the pair center-of-mass frame.
Also shown are $pp$ correlations for the same source sizes.
Clearly, $\Lambda p$ correlations are more sensitive than $pp$
correlations for determining larger source sizes. 
Despite the fact that the $pp$ scattering length is considerably longer, 
Coulomb effects obscure the sensitivity of $pp$ correlation for sources 
larger than approximately 6~fm. 
Thus, even though statistics for $\Lambda p$ correlations are reduced 
compared to $pp$ statistics, they may provide a more accurate 
determination of the source size under some circumstances. 
Correlations involving neutrons, such as $nn$ or $pn$, 
are also free of a relative Coulomb interaction, and given the large 
$nn$ and $pn$ scattering lengths, provide sensitive correlations. 
However, neutrons are notoriously difficult to measure, 
whereas lambdas can be measured with a charged-particle detector 
through its $\pi^-p$ decay channel.
Another candidate for source size measurement is 
deuteron coalescence~\cite{llope}. 
Statistically, coalescence should provide the most accurate 
determination of source parameters of all baryonic probes as
coalescence does not suffer from Coulomb effects, 
and employs the entire strength of a bound state rather than 
the fraction represented by a rise in the scattering phase shift.

For illustration, we have used the same source size for both 
lambdas and protons, and assumed thermal momentum distributions.
To eliminate the Lorentz factor effect and also for computational reasons, 
we have used thermal temperature $T$=3~MeV. 
However, the Lorentz factor effect is small due to 
the large lambda and proton masses. 
For instance, the difference in the correlation functions 
between $T$=3 and 300~MeV is less than 5\%.
For the interest of experimental feasibility of $\Lambda p$ 
correlation measurements, for $T$=300~MeV, a total of 20 million pairs
results in, from pure phase space population,
10 pairs in 0$<$$k$$<$5~MeV/$c$ (hence 30\% statistical uncertainty),
70 pairs in 5$<$$k$$<$10~MeV/$c$ (12\%), 
800 pairs in 25$<$$k$$<$30~MeV/$c$ (4\%),
and 7000 pairs in 0$<$$k$$<$50~MeV/$c$.

An experimental correlation function often resorts to a mixed-event 
technique for uncorrelated pairs. 
Since both the height and the width of a $\Lambda p$ correlation 
function are sensitive to the source size, 
it is important to normalize the correlation function properly. 
Since lambda and proton exhibit no correlation at large $k$, 
an experimental $\Lambda p$ correlation function may be normalized 
to unity at large $k$.

\begin{figure}
\centerline{\epsfxsize=0.45\textwidth \epsfbox{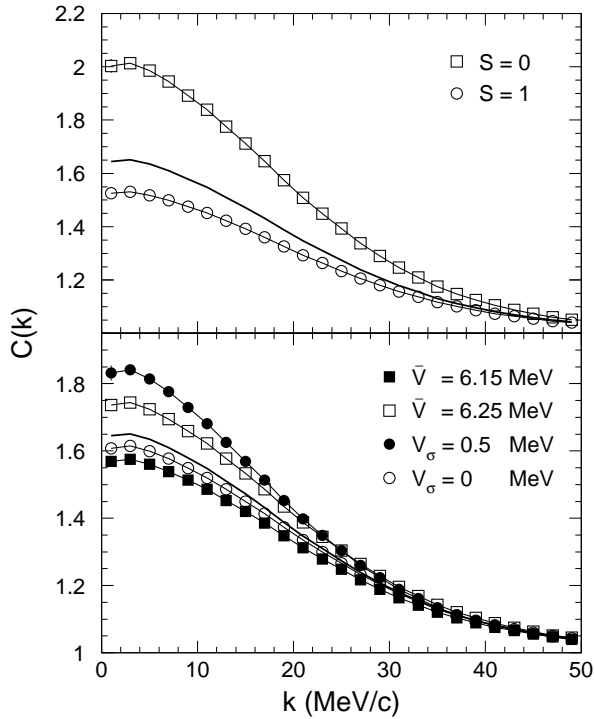}}
\caption{Upper panel: the spin decomposition of the $\Lambda p$ 
correlation function for a $R_g$=4~fm source, 
with the spin-averaged result represented by the thick solid line. 
Lower panel: the sensitivity of the correlation function to 
the uncertainties in the potential parameters. 
Default values ($\bar{V}$=6.2~MeV, $V_\sigma$=0.25~MeV) are
represented by the thick solid line. 
Changing $\bar{V}$ yields the results represented by the squares; 
changing $V_{\sigma}$ yields the results represented by the circles. 
The uncertainties in the parameters translate to a $\pm 0.5$~fm 
uncertainty in the extracted source size.}
\label{fig2_spin}
\end{figure}

The $\Lambda p$ correlation function is one fourth 
spin singlet ($S$=0) and three fourths spin triplet ($S$=1).
To illustrate the spin dependence, correlation functions are presented
for a $R_g$=4~fm source in the upper panel of Fig.~\ref{fig2_spin}, 
separately for $S$=0 and $S$=1 pairs. 
If $V_{\sigma}$ in Eq.~(\ref{eq:urbana}) were zero, 
the two contributions would be identical. 
As $V_{\sigma}$ is not well understood, $0<V_\sigma<0.5$~MeV~\cite{urbana},
measuring the spin dependence of the correlation function could 
in principle determine $V_\sigma$.

The correlation function is largely determined by the scattering length
and effective range of the potential~\cite{lyuboshits}.
The scattering length and effective range corresponding to the 
potential~(\ref{eq:urbana}) ($\bar{V}$=6.2~MeV, $V_\sigma$=0.25~MeV)
are $-2.88$~fm and $2.92$~fm for the spin singlet, 
and $-1.66$~fm and $3.78$~fm for the spin triplet, respectively.
They are in reasonable agreement with those from Refs.~\cite{urbana,nagels}.
Using these values and an analytical approximation similar for 
neutron-proton correlations~\cite{lyuboshits}, we obtain 
$\Lambda p$ correlation functions that are consistent with the results 
in the upper panel of Fig.~\ref{fig2_spin}.

The sensitivity of the spin-averaged correlation function to 
the parameters $\bar{V}$ and $V_\sigma$ is illustrated 
in the lower panel of Fig.~\ref{fig2_spin}. 
Using the stated uncertainties~\cite{urbana}, 
$\bar{V} = 6.2 \pm 0.05$~MeV and $V_{\sigma} = 0.25 \pm 0.25$~MeV, 
the correlation function for a $R_g$=4~fm source is shown for 
a range of parameters. 
The results suggest that the uncertainties in the potential parameters 
translate into an approximately $\pm 0.5$~fm uncertainty in the 
source size extracted from the correlation function at $k$$<$25~MeV/$c$, 
while the large $k$ tail has better constrain on the source size.

\begin{figure}
\centerline{\epsfxsize=0.45\textwidth \epsfbox{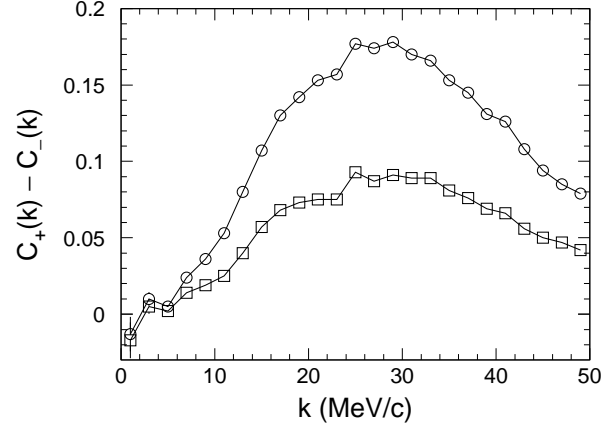}}
\caption{The difference in the $\Lambda p$ correlation functions 
for positive and negative values of $k_z$ is shown assuming that 
both lambdas and protons are characterized by 4~fm sources, 
but are separated by 1.5~fm (squares) or 3~fm (circles) in $z$. 
Such a separation might result 
if the lambdas would escape earlier than the protons. 
Small-scale fluctuations are due to fluctuations 
in the Monte Carlo procedure.}
\label{fig3_lednicky}
\end{figure}

Recently, Lednicky {\em et~al.}~\cite{lednicky} have shown that 
correlations of non-identical particles can provide information 
revealing whether the particles are emitted simultaneously. 
If the lambdas are emitted before the protons in such a way that 
the probability cloud describing the protons lags that for the
lambdas of the same velocity, the correlation function then 
depends on the sign of the relative momentum in the direction 
defined by that of the displacement of the lambda and proton clouds. 
An example is illustrated in Fig.~\ref{fig3_lednicky}, 
where both sources are assumed to be characterized by a size of 4~fm, 
but are separated by $\Delta z$=1.5 fm or 3.0 fm. 
In Fig.~\ref{fig3_lednicky} the difference of the correlation functions,
$C_+(k)-C_-(k)$, is plotted against $k$ where the transverse component 
of $k$ is required to be less than 10 MeV/$c$. 
Here, $C_+$ refers to the correlation function constructed 
with the requirement that $k_z>0$, 
while $C_-$ is constructed with the opposite constraint.

A displacement $\Delta z$ can result from a displacement in time, 
$\Delta \tau=\Delta z/v$, where $v$ is the velocity of the $\Lambda p$ pair. 
The displacement direction, $\hat{z}$, is then defined by the direction 
of $v$, or that of the pair momentum relative to the source. 
Thus, if $\Lambda p$ correlations can be measured with an accuracy of a few 
percent at $k$$\sim$30~MeV/$c$, the conjecture that strange and non-strange 
baryons are emitted simultaneously can be addressed quantitatively.

One should remember that the residual interaction of the proton with 
the Coulomb field of the nuclear sources might distort the result. 
Unlike $pp$ correlations, where both particles feel the identical force, 
only the proton experiences the Coulomb field. 
This issue has been previously considered in the context of 
$pn$ correlations where the distortion was shown to be small 
for fast-moving pairs~\cite{erazmus}. 
This effect should be smaller at RHIC where the excess charge at midrapidity 
is expected to be smaller than observed at SPS and AGS energies.

Finally, we make a brief note regarding $\Lambda\Lambda$ 
and $\bar{\Lambda} p$ correlations.
In the view of our $\Lambda p$ results, we expect that the 
$\Lambda\Lambda$ strong interaction would also give sizeable correlations.
These correlations are of great interest because of the predicted 
existence of a $\Lambda\Lambda$-like particle~\cite{jaffe}.
In conjunction with $\Lambda p$, $\bar{\Lambda}p$ correlations may reveal 
valuable information on low energy $\bar{\Lambda}p$ annihilation
cross sections which are presently unknown but indispensable in modeling 
certain aspects of heavy-ion collisions~\cite{wang}.

In summary, $\Lambda p$ correlations may provide a useful characterization 
of the space-time structure of relativistic heavy ion collisions 
should one be able to gather sufficient statistics. 
The lack of a relative Coulomb interaction allows the strong interaction 
to produce a large peak in the correlation function even for large sources,
to which $pp$ correlation loses its sensitivity.
Furthermore, by binning according to the sign of the projected relative 
momentum, one might address the question of whether lambdas and protons 
are emitted simultaneously. 
However, the interpretation of the correlation function could benefit 
from a more precise parameterization of the $\Lambda p$ interaction.

\acknowledgements{
F.W. acknowledges Drs. V.~Koch, R.~Lednicky, J.C. Peng, 
A.M.~Poskanzer and N.~Xu for useful discussions. 
This work was supported by the U.S. Department of Energy 
under contract DE-AC03-76SF00098 and 
the National Science Foundation under grant PHY-96-05207.}


\end{document}